\documentclass[global,twocolumn]{svjour}
\usepackage{graphics}
\usepackage{amsmath}

\usepackage{todonotes}
\usepackage{draftwatermark}
\SetWatermarkScale{1.5}
\SetWatermarkLightness{.85}
\SetWatermarkText{\shortstack{}}

\journalname{myjournal}
\begin{document}
\title{The trap design of PENTATRAP}

\author{C. Roux\inst{1,2} \and Ch. B\"ohm \inst{1,2,3} \and A. D\"orr \inst{1,2} \and S. Eliseev \inst{1} \and S. George \inst{1}\thanks{\emph{Present address:} National Superconducting Cyclotron Laboratory, MSU, East Lansing, MI, 48824-1321, USA} \and M. Goncharov \inst{1,2} \and Yu. Novikov \inst{3,4} \and J. Repp \inst{1,2} \and S. Sturm \inst{1,5} \and S. Ulmer \inst{1,2,5}\thanks{\emph{Present address:} RIKEN Advanced Science Institute, Hirosawa, Wako, Saitama 351-0198, Japan} \and K. Blaum \inst{1,2} 
%
}                     
\offprints{christian.roux@mpi-hd.mpg.de}          
\institute{
Max-Planck-Institut f\"ur Kernphysik, 69117 Heidelberg, Germany \and 
Fakult\"at f\"ur Physik und Astronomie, Ruprecht-Karls-Universit\"at, 69120 Heidelberg, Germany \and
Extreme Matter Institut EMMI, Helmholtz Gemeinschaft, 64291 Darmstadt, Germany \and
St. Petersburg Nuclear Physics Institute, 188300 Gatchina, Russia \and
Institut f\"ur Physik, Johannes Gutenberg-Universit\"at, 55099 Mainz, Germany
}
\date{Received: date / Revised version: date}
%
\maketitle
øø
\begin{abstract}
A novel Penning trap tower consisting of five compensated cylindrical Penning traps is developed for the PENTATRAP mass spectrometer at the Max-Planck-Institut f\"ur Kernphysik in Heidelberg, Germany.
An analytical expression for the electrostatic potential inside the trap tower is derived to calculate standard Penning trap properties like the compensation of anharmonicities and an orthogonal geometry of the trap electrodes. Since the PENTATRAP project described in the preceding article aims for ultra high-precision mass-ratio measurements of highly charged ions up to uranium, systematic effects for highly charged ions  inside the trap tower are considered for the design process as well. Finally, a limit due to remaining anharmonic shifts at large amplitudes is estimated for the resulting geometry, which is important for  phase-sensitive measurements of the reduced cyclotron frequency of the ions.
\end{abstract}
\section{Introduction}
\label{intro}
The mass of atoms and nuclei are of great interest in many fields of modern physics
due to its inherent connection with the atomic and nuclear binding energies. The list of applications of precise mass or mass-ratio values ranges from tests of nuclear and atomic structure, determination of fundamental constants and main features of the Standard Model for elementary particles to measurements related to neutrino physics \cite{Blaum:10:ContPhys}. This general need for very accurate mass values has triggered an ongoing technical development in mass spectrometers. Penning traps have proven to be the ideal devices for the determination of atomic masses, capable of reaching extremely high accuracy down to $\delta m/m \simeq 10^{-11}$. Worldwide nearly 20 Penning trap mass spectrometers are in operation or under construction \cite{Blaum:06:PhysRep}.\\
With a destructive time-of-flight cyclotron frequency detection method, the most precise mass values  reached so far for stable atoms have a relative accuracy of $\delta m/m \sim 6\cdot10^{-10}$ \cite{Bergstroem:03:EurPhyJD,Nagy:06:PhysRevLett}. Even for extremely short-lived species (down to $t_{1/2}=8.8$ ms) or very low production rates of less than one atom per second in the case of heavy and superheavy ions, relative accuracies down to $5\cdot10^{-8}$ have been shown \cite{Smith:08:PhysRevLett,Block:10:Nature}. For light and stable atoms and with a non-destructive image current detection method, results with an even more striking accuracy below the $10^{-11}$ barrier have been achieved, allowing, e.g., for the most stringent direct test of Einstein's energy-mass-relation $E=mc^2$ \cite{Rainville:05:Nature} and the most precise mass measurement of $^4$He \cite{VanDyck:04:PhysRevLett}.\\
This outstanding accuracy in Penning trap mass spectrometry can be reached, since the measurement of the mass $m$ of an ion with charge $q$ is done via a determination of the cyclotron frequency
\begin{eqnarray}
\nu_c=\frac{1}{2\pi}\omega_c=\frac{1}{2\pi}\frac{q}{m}B,
\end{eqnarray}
where $B$ is the magnetic field strength of the trap's magnet.
In Penning traps, the free cyclotron frequency $\nu_c$ can be determined by application of the Brown-Gabrielse invariance theorem \cite{Brown:82:PhyRevA}
\begin{eqnarray}
\nu_c^2=\nu_+^2+\nu_z^2+\nu_-^2,\label{eq:invariancetheorem}
\end{eqnarray}
where $\nu_+$, $\nu_z$ and $\nu_-$ are the eigenfrequencies of the charged particle in the presence of the trap's magnetic and electrostatic field. Eq. (\ref{eq:invariancetheorem}) is immune to small misalignments between the electrostatic axis of the trap and the magnetic field and also to small ellipticities of the trap electrodes. An exhaustive review of the physics of single charged particles in a Penning trap can be found in \cite{Brown:86:RevModPhys}.\\
The accuracy of measurements of atomic masses or mass-ratios is limited by a variety of systematic and statistical effects. Examples are magnetic or electric trapping field imperfections and their temporal stability, eigenfrequency shifts arising from the detection system, image charge effects or relativistic shifts of the ion's mass themselves (see e.g. \cite{Kellerbauer:02:EurPhyJD,VanDyck:06:IntJMS}).
The most crucial effect setting limits to the precision of the measurements is expected to be the temporal stability of the magnetic field. There are several concepts in current state-of-the-art experiments to minimize this influence. A very obvious approach has led to the development of an advanced stabilization system for superconducting magnets \cite{VanDyck:99:RevScI}, where relative drifts of $\delta B/B\approx2\cdot 10^{-12}/\text{h}$ were realized.
The temporal stability is even more crucial, since high-precision measurements are always performed in a relative way, which means that only frequency ratios between two ion species are measured. Therefore, not only the measurement itself, but also the loading and preparation of the two species has to be performed very quickly. Taking this to the limit, a very elegant method has been developed for low charged ions, where both ions are prepared on a common magnetron radius and measured at the same time \cite{Rainville:04:Science}. Thus, temporal drifts and noise act on both ions in the same way and cancel out in the frequency ratio determination.\\
For highly charged ions it is not possible to measure the mass ratio of two ions with high accuracy simultaneously in the same trap, due to the strong Coulomb interaction and the resulting systematic shift. Thus, the ions have to be measured either alternately in one trap, or simultaneously in two different traps. 
Our project PENTATRAP \cite{Repp:11:ApplPhysB}, which aims for measurements of mass ratios of highly charged ions with an accuracy of $<10^{-11}$, will enable both options. For this purpose, five identical cylindrical Penning traps will be used to provide the possibility of flexible measurement schemes and fast exchanges of ions between traps.
The latter option, where the cyclotron frequencies of the two ions are measured simultaneously in adjacent traps is certainly favorable, since it suppresses statistical variations in the cyclotron frequency ratio due to temporal fluctuations of the B-field.
Additionally, one trap can be used to store an arbitrary ion and permanently monitor its cyclotron frequency to detect  temporal fluctuations of the magnetic field or to serve as a reference for the stabilization of the trap potential. The possible measurement schemes of this trap setup and ion candidates are described in detail in the preceding article by J. Repp et al. \cite{Repp:11:ApplPhysB}.
Besides temporal fluctuations of the magnetic field, which are taken care of by the measurement scheme, numerous other systematic effects have to be taken into account and intensive studies of the complete measurement system have to be performed.\\
The subject of this paper is the design of the five-Penning-trap tower, which implies the calculation of the ideal orthogonal geometry of the compensated traps. Furthermore, systematic effects for highly charged ions inside the traps like image charge shifts and the Coulomb repulsion between two ions in adjacent traps are considered for the design process. Finally, machining imperfections are included and limits for measurements at large amplitudes are estimated, which is of special importance for the measurement of the reduced cyclotron frequency.

\section{Cylindrical Penning traps}\label{sec:CylPenn}
The natural geometry of a Penning trap utilizes hyperboloidly shaped electrodes to generate a quadrupolar potential in the trap center \cite{Brown:86:RevModPhys}.
In our case cylindrical trap electrodes with open endcaps \cite{Gabrielse:84:IntJMS} will be used. This trap geometry enables an easy injection of the particles or exchange between different traps.  Additionally, cylindrical trap electrodes can be machined with higher precision and the resulting trap potential can be calculated analytically, which is crucial for the design process. 
Therefore, this geometry has become very popular in recent years and is widely used at various Penning trap experiments like high-precision mass measurements, see e.g. \cite{Blaum:10:ContPhys,Blaum:06:PhysRep} and references therein, \textit{g}-factor measurements of the bound electron and the proton \cite{Blaum:09:JPhysB} as well as the free electron and the positron \cite{Hanneke:08:PhysRevLett}, cooling of highly charged ions \cite{Kluge:08:AdvQuanChem} or the capture of anti-protons \cite{Gabrielse:86:PhysRevLett} and for the production of anti-hydrogen \cite{Gabrielse:02:PhysRevLett,Amoretti:02:Nature}.\\
If very high frequency resolution has to be achieved, it is essential that the electrostatic potential is harmonic along the trap axis to ensure that the axial frequency is independent of the energy of the ion. Therefore, correction electrodes are added to compensate, e.g., for machining imperfections. This was first done for hyperboloidal traps in \cite{VanDyck:76:ApplPhysLett} and also applied to cylindrical traps in \cite{Gabrielse:84:IntJMS}. However, an additional correction voltage generally changes the curvature of the harmonic potential and therefore the ion's axial frequency, which is very undesirable for practical use. Thus, a special geometry can be found where the harmonic part of the potential is independent of the applied correction voltage and stays constant while tuning out anharmonicities \cite{Gabrielse:84:IntJMS}.

\subsection{Potential of a cylindrical trap tower}\label{sec:CylPennPotential}
An analytical expression for the potential of cylindrical trap geometries can be derived in a standard way from the Laplace-equation with azimuthal symmetry 
\begin{eqnarray}
\nabla^2\phi(\rho,z)=\frac{1}{\rho}\frac{\partial}{\partial \rho}\left(\rho \cdot\frac{\partial \phi}{\partial \rho}\right)+\frac{\partial^2\phi}{\partial z^2}=0. 
\end{eqnarray}
For a potential invariant under $z \mapsto -z$ transformation, the general solution is given by \cite{jacksonPot}
\begin{eqnarray}
\phi(\rho,z)=\int_{-\infty}^{\infty}\limits dk A(k) I_0(k\rho)\cos(kz).
\end{eqnarray}
Here, $k$ results from the separation of variables and $I_0$ is the modified Bessel function of zero order and first kind. The coefficient $A(k)$ is defined by the boundary conditions, which are given by the geometry of the trap electrodes and the voltages applied to them. Unfortunately, this integral cannot be calculated analytically.
Assuming closed and grounded endcaps, the potential has to vanish at the ends of the trap and the integral can be represented by an infinite sum, given by
 \begin{eqnarray}
\phi(\rho,z)=\sum_{n=-\infty}^{\infty}\limits A(k_n) I_0(k_n\rho)\cos(k_nz),
\end{eqnarray}
where $k$ is replaced by the discrete $k_n=\frac{n\pi}{l}$. Here, $l$ is the total length of the trap tower. As discussed  for a single open-endcap trap in \cite{Gabrielse:89:IntJMS}, the suitability of this approximation for open-endcap traps depends on the ratio of the inner radius and the length of the grounded endcaps. It gives results valid to better than $1~\%$ in all electrostatic coefficients if the length of the endcaps is three times the inner radius. In our case, this ratio is even higher which therefore leads to much more precise predictions about the electrostatic coefficients, as will be shown later.\\
The coefficients $A(k_n)$ can be obtained by using the orthogonality relation of the cosine function and integrating over $z$ along the inner surface of the electrodes at $\rho = a$. The potential at $\rho=a$ at the small gaps between the electrodes can be approximated in axial direction by a linear interpolation between the values of the two adjacent electrodes. This leads to a negligible error if the length $d$ of the gaps is much shorter than the length scales of the electrodes \cite{Verdu:08:NJP}. In doing so, the following expression results for a trap with grounded endcaps and $m=(2\kappa+1)$ electrodes:
\begin{align}
\phi (\rho, z)&=\sum_{n=1,odd}^{\infty}A(k_n)I_0\left(k_n \rho\right)\cos\left(k_n z\right)\label{eq:FallenPotential} \\ 
A(k_n)&=\frac{4}{l}\frac{1}{k_n^2 d}\sum_{i=1}^{\kappa}(U_{i}-U_{i-1})\nonumber\\
&\times\frac{\cos\left(k_n z_{2i}\right)-\cos\left(k_nz_{2i-1}\right)}{I_0\left(k_n a\right)}.
\end{align}
\begin{figure}
\centering
\includegraphics[width=0.45\textwidth]{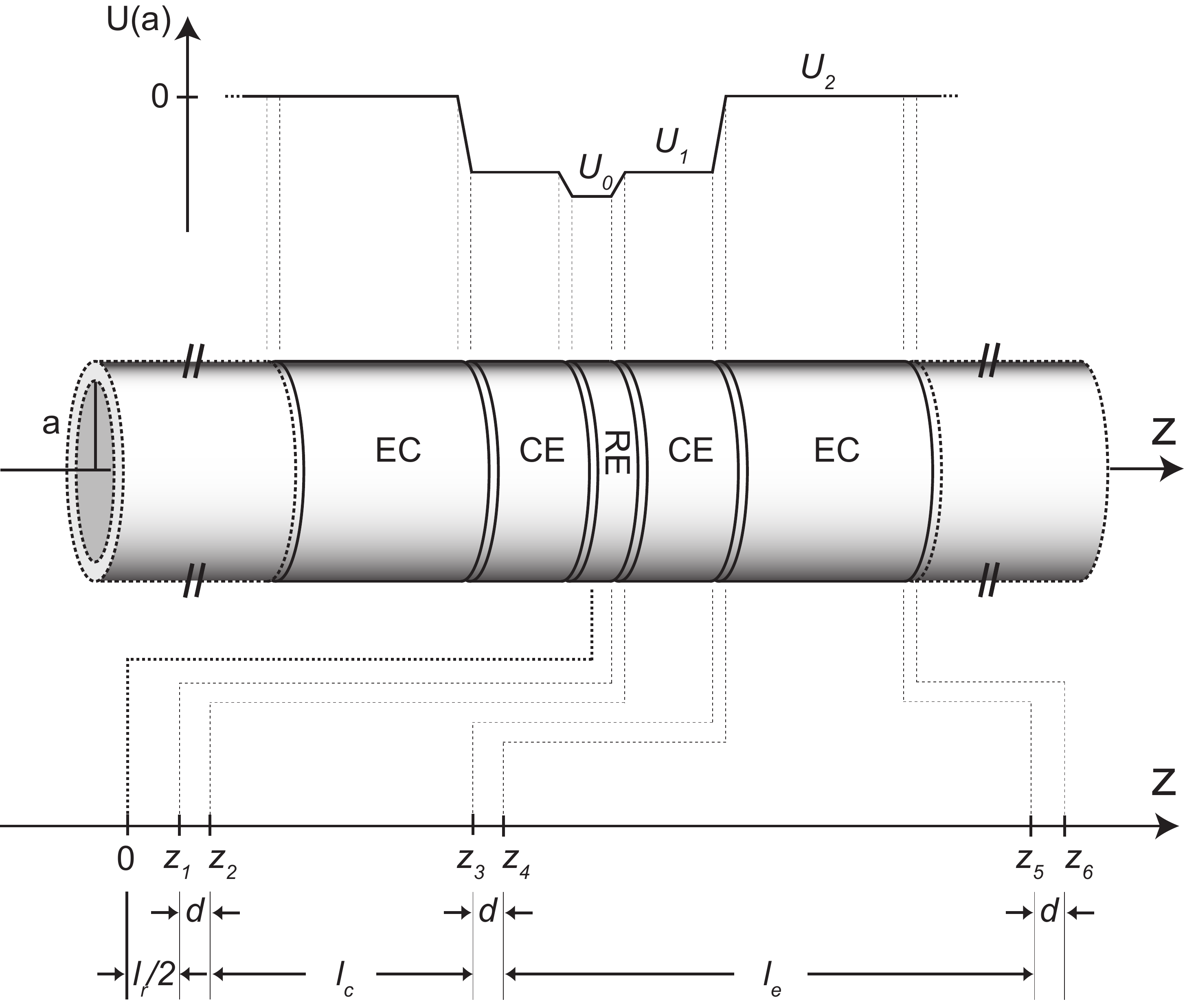}
\caption{Illustration of the a five-electrode cylindrical trap inside a trap tower. The inner radius is given by $a$. The lengths of the trap electrodes are marked with $l_r=2 z_1$ for the central ring electrode RE, $l_c=z_3-z_2$ for the correction electrodes CE and $l_e=z_5-z_4$ for the endcaps EC. The length of the gaps separating the electrodes is given by $d$. The potentials at the inner surface of the electrodes are given by $U_i$ with a linear approximation at the gaps between the electrodes.}
\label{fig:fallenskizze}
\end{figure}
As shown in Fig.~\ref{fig:fallenskizze}, $U_i$ are the voltages applied to the electrodes, $z_{2i}$ and $z_{2i-1}$ specify the ends of the individual electrodes and $a$ is the inner radius. In our case, every trap consists of five electrodes, in particular of a central ring electrode RE with length $l_r=2 z_1$, two correction electrodes CE with length $l_c=z_3-z_2$ and two grounded endcaps EC with length $l_e=z_5-z_4$.\\
The potential in Eq.~(\ref{eq:FallenPotential}) allows one to deduce analytical expressions for the static electric field coefficients. 
For hyperboloidal traps, extensive studies of these coefficients and the influence on the ion's motion can be found in \cite{Brown:86:RevModPhys,Gabrielse:83:PhysRevA}. In \cite{Gabrielse:84:IntJMS} these results are transferred to cylindrical traps. In all of these works the potential is expanded in spherical coordinates in terms of Legendre polynomials. Additionally, in order to make the expansion coefficients dimensionless, the polynomials are multiplied by powers of the small ratio $r/d_0$, where $r$ is the radial coordinate and $d_0$ a characteristic  trap parameter \cite{Brown:86:RevModPhys}.\\
In our work we follow another ansatz as shown in \cite{Verdu:08:NJP}, where the trapping potential is Taylor expanded in cylindrical coordinates $[\rho, \varphi, z]$ at the center of the trap,
\begin{eqnarray}
\phi(\rho,z)&=&\sum_{j=0}^{\infty}\limits\sum_{i=0}^{j}\limits\frac{1}{i! (j-i)!}\left.\frac{\partial^j\phi}{\partial \rho^i\partial z^{j-i}}\right|_{(0,0)}\rho^iz^{j-i}\\
&\equiv& U_0\sum_{j=0}^{\infty}\limits\sum_{i=0}^{j}\limits c_{i,j-i}\rho^iz^{j-i}.\label{eq:TaylorPotential}
\end{eqnarray}
The coefficients $c_{i,j-i}=\frac{1}{U_0}\frac{1}{i! (j-i)!}\left.\frac{\partial^j\phi}{\partial \rho^i\partial z^{j-i}}\right|_{(0,0)}$ are normalized to the ring voltage $U_0$. Due to the symmetry of reflection in $z$, all terms with odd order have to vanish to fulfill the boundary conditions. Therefore, both $i$ and $j$ have to be even numbers. 
In the following, we will use only the pure axial coefficients 
\begin{eqnarray}
c_j\equiv c_{0,j}=\frac{1}{U_0}\left.\frac{1}{j!}\frac{\partial^j \phi(\rho,z)}{\partial z^j}\right|_{(0,0)}\label{eq:Ckoeffizienten}
\end{eqnarray}
to analyze the potential. Due to $\nabla^2 \phi(\rho, z) = 0$, the knowledge of the potential on the axis also determines the radial behavior.
These Taylor coefficients are only functions of the geometry and the applied voltages and can, therefore, directly be used to analyze the harmonicity of the trap geometry. By comparison of the coefficients of same order $j$ to the Legendre expansion used in \cite{Gabrielse:84:IntJMS}, it can be shown that the two kind of coefficients simply differ by a factor of $1/(2d_0^j)$. Therefore, all results in \cite{Gabrielse:84:IntJMS} concerning the ion's motion can be used and are shortly summarized below in terms of the coefficients defined in Eq.~(\ref{eq:Ckoeffizienten}).\\
Certainly, the lowest order coefficients in Eq.~(\ref{eq:TaylorPotential}) are the most important, since the axial amplitude of the ion is much smaller than the trap dimensions. For an ideal harmonic trap, the axial oscillation frequency is given by 
\begin{eqnarray}
\omega_z=\sqrt{\frac{qU_0}{m}2c_2}\label{eq:AxialFreqIdeal}.
\end{eqnarray}
The higher order coefficients determine the anharmonicity of the trapping potential. They result in an amplitude dependent shift in the axial oscillation frequency. For the leading anharmonic coefficients $c_4$ and $c_6$ the shift in the axial frequency $\omega_z$ is given by 
\begin{eqnarray}
\frac{\Delta \omega_z}{\omega_z}=  \frac{3}{4}\left(\frac{c_4}{c_2^2}+\frac{5}{4}\frac{c_6}{c_2^3}\frac{E_z}{qU_0}\right)\frac{E_z}{qU_0},
\label{eq:AnharmonicShiftTaylor}
\end{eqnarray}
where $E_z$ is the axial excitation energy of the ion.
In a five-electrode cylindrical trap, the potential can be tuned to minimize this anharmonic shift by adjusting the correction voltage $U_1\equiv U_c$ (see Fig.~\ref{fig:fallenskizze}). Since the potential in Eq.~(\ref{eq:FallenPotential}) is linear in the applied voltages, the Taylor coefficients of this potential can be written as \cite{Verdu:08:NJP}
\begin{eqnarray}
c_j:=e_j+ T d_j,\label{eq:CED}
\end{eqnarray}
where $T\equiv U_c/U_0$ is the "tuning ratio" between the ring voltage $U_0$ and the correction voltage $U_c$. The $e_j$ coefficients determine the part which is independent of the correction voltage, while the $d_j$ coefficients defined by $d_j\equiv\partial c_j/\partial T$ contain the influence of $U_c$. If \textit{T} is chosen to be $T=\left.T\right|_{c_4=0}=-e_4/d_4$, the leading anharmonic coefficient $c_4$ cancels out. 
Due to the dependence of $c_2$ on $U_c$, tuning of the anharmonic coefficients by adjustment of $U_c$ is accompanied with an undesirable change of $c_2$ and, therefore, also of the axial frequency in Eq.~(\ref{eq:AxialFreqIdeal}). To avoid this, a geometry has to be found in which $d_2$ and, therefore, the dependence of the axial frequency on the correction voltage is minimized. This trap geometry is called \textit{orthogonal}.

\subsection{Ion-ion interaction}\label{sec:IonIon}

Since we want to store at least three ions in adjacent traps in the final measurement process of our high-precision Penning trap mass spectrometer, an important issue is the Coulomb repulsion and the possibly resulting shift of the free cyclotron frequency.\\
To analyze this effect, the additional ion-ion potential $\phi_{ii}(\vec{r},\vec{r}')$ induced by a charge at $\vec{r}'$ has to be calculated at the position $\vec{r}$ of the ion-of-interest. 
Thereby, for our case the distance between two trap centers ($\Delta z \sim 10$~mm) is much larger than the typical axial oscillation amplitude of the ions ($\sim 10~\mu$m) and all dynamical aspects can be neglected.\\
The analysis can be performend by expanding the ion-ion potential in a Taylor series with cylindrical coordinates $[\rho, \varphi, z]$, in a way similar to the expansion of the trap potential in the previous section:
\begin{eqnarray}
&~&\phi_{ii}^{taylor}(\rho,z,\rho',z')=\nonumber\\
&=&\sum_{n=0}^{\infty}\limits\sum_{m=0}^{n}\limits\frac{1}{m! (n-m)!}\left.\frac{\partial^n\phi_{ii}(\rho,z,\rho',z')}{\partial \rho^m\partial z^{n-m}}\right|_{(0,0)}\rho^mz^{n-m}\nonumber\\
&\equiv& \sum_{n=0}^{\infty}\limits\sum_{m=0}^{n}\limits g_{m,n-m}\rho^mz^{n-m}.\label{eq:TaylorPotentialIonIon}
\end{eqnarray}
In contrast to the trap potential, for this interaction there is no symmetry with respect to the $z$-axis at the position of the ion-of-interest. Nevertheless, by inserting Eq.~(\ref{eq:TaylorPotentialIonIon}) into the Laplace equation it can be shown that the following relation of the Taylor coefficients holds for potentials with azimuthal symmetry:
\begin{align}
\phi_{ii}^{taylor}=&~g_{0,0}+g_{1,0}\cdot\rho+g_{0,1}\cdot z+\nonumber\\
&+g_{2,0}\cdot \rho^2+g_{1,1}\cdot\rho z+g_{0,2}\cdot z^2+\\
&+{\mathcal O}(\rho^3,z^3),\nonumber\\
\text{with}~~
g_{1,0}&=0,~~g_{1,1}=0~~\text{and}~~g_{2,0}=-\frac{1}{2}g_{0,2}.
\end{align}
Hence, in second order this potential has exactly the same shape as the quadrupolar trap potential \cite{Gabrielse:84:IntJMS}. In other words, in second order the ion-ion potential acts as a small offset on the trapping voltage, which exactly cancels in the determination of the free cyclotron frequency $\nu_c$ \cite{Brown:86:RevModPhys}.\\
The remaining effect is a repulsion of the ions induced by $g_{0,1}$ resulting in a shift of the axial equilibrium position. 
The new minimum can be calculated from the superposition of the ion-ion potential and the harmonic part of the trap potential $\phi(r,z)=U_0c_2(z^2-\rho^2/2)$ defined in Eq.~(\ref{eq:TaylorPotential}). The minimum is given by
\begin{eqnarray}
\tilde{z_0}=-\frac{g_{0,1}}{2(U_0 c_2+g_{0,2})},
\end{eqnarray}
where $C_2\equiv U_0c_2$ can be determined by Eq.~(\ref{eq:AxialFreqIdeal}) for every $q/m$ ratio.
The coefficients $g_{0,1}$ and $g_{0,2}$ can be calculated by the potential of a point-charge in a conducting cylinder with inner radius $a$ and length $L$ given by \cite{jackson}
\begin{align}
\phi_{ii}(\vec{r},\vec{r}')&=\frac{q}{\pi \epsilon_0 a}
\sum_{m=-\infty}^{\infty} e^{i m (\varphi-\varphi ')} \\
&\times  \sum_{n=1}^{\infty}
\frac{J_m\left(\frac{\rho}{a}\cdot x_{mn}\right)J_m\left(\frac{\rho'}{a}\cdot x_{mn}\right)}{x_{mn}J^2_{m+1}(x_{mn})
\sinh\left(\frac{L}{a}\cdot x_{mn}\right)}\nonumber\\
&\times \sinh\left(\frac{z_<}{a}\cdot x_{mn}\right)\sinh\left(\frac{L-z_>}{a}\cdot x_{mn}\right)
.\nonumber
\end{align}
Here, $x_{mn}$ is the \textit{n}-th root of the Bessel function $J_m$ of order $m$ and $z_{>/<}$ are the larger or smaller value of $z$ and $z'$.\\
Certainly the largest shift is expected for a shallow trapping potential corresponding to a large $q/m$ ratio. 
On the other hand, the Coulomb repulsion in the trap tower is strongly suppressed compared to free space due to the shielding of the surrounding trap electrodes. \\
Since the homogeneous part of our magnet is limited to $120$~mm, the maximum length of one trap is given by $24$~mm. It turns out that at this distance and for a large inner radius of $a=5$~mm, this shift in the origin of oscillation is neglibigle for all experimental situations. Even in the extreme case of a proton as ion-of-interest and a charge of $q=81+$ in the adjacent trap, the shift  is in the order of $\tilde{z_0}\sim 10^{-12}$~m. Therefore, ion-ion interaction can be safely ignored for our trap tower.

\subsection{Image charge effects}
\begin{figure}
\centering
\includegraphics[width=0.45\textwidth]{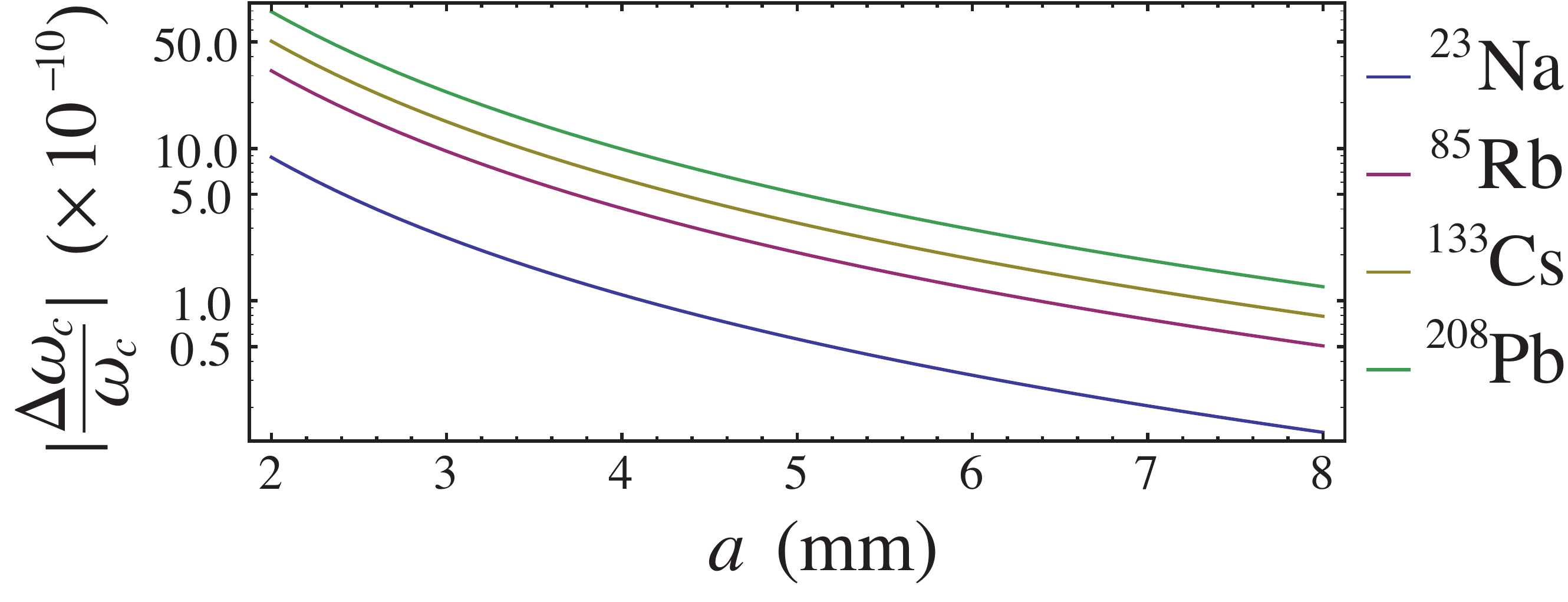}
\caption{Relative shift of the free cyclotron frequency due to image charges induced in the trap electrodes as a function of the inner radius $a$ of the trap electrodes. Parameters used for the calculation are $B=7$~T and $\omega_z=2\pi\cdot 600$~kHz. For the calculation, the $q/m$ ratio was chosen sufficiently high such that Eq.~(\ref{eq:imagechargeshift}) becomes independent of the charge state $q$ of the ions.}
\label{fig:spiegelladungen}
\end{figure}

As pointed out in \cite{VanDyck:89:PhyRevA,Porto:01:PhysRevA} for hyperboloidal Penning traps, the shift of the normal-mode frequencies in a Penning trap due to the electric field of image charges induced in the trap electrodes is not negligible for highly charged ions.
In the case of cylindrical trap geometries, the resulting electric field and, thus, the force on the ion can be calculated analytically by solving the Laplace-equation as shown in \cite{Haeffner:00:phd}. The shift of the radial frequencies is given as a function of the trap radius $a$
\begin{eqnarray}
\Delta\omega_{\pm}\approx \mp\frac{q^2}{4 \pi \epsilon_0 m a^3 \omega_c},
\end{eqnarray}
whereas the axial frequency is not shifted, since the image charge electric field is translationally invariant in a cylindrical trap.
The influence on the determination of the free cyclotron frequency can again be calculated by using the invariance theorem \cite{Brown:82:PhyRevA}:
\begin{eqnarray}
\frac{\Delta \omega_c}{\omega_c}\approx \left(-\frac{\omega_+}{\omega_c}+\frac{\omega_-}{\omega_c}\right)\frac{q^2}{4 \pi \epsilon_0 m a_0^3 \omega_c^2}.\label{eq:imagechargeshift}
\end{eqnarray}
The resulting shift is shown in Fig.~\ref{fig:spiegelladungen} for different masses. Note that for typical experimental conditions with $\omega_+\gg\omega_z\gg\omega_-$,  Eq.~(\ref{eq:imagechargeshift}) is independent of the charge $q$ of the ion.
From this point of view it is convenient to have an inner radius as large as possible. At $a=5$~mm, the shift in the free cyclotron frequency is still at $\sim 5\cdot10^{-10}$ for heavy ions like $^{208}$Pb.\\ 
At charge states $q>20+$, the relative shift per mass unit $u$ is below $2.5\cdot10^{-12}$ for masses between $A=100$~u and $A=250$~u. Thus, if reference ion and ion-of-interest have nearly the same mass, the image charge effect on the ratio becomes negligible.
As mentioned before, main interests of our project are neutrino related mass-ratio measurements or measurements of binding energies of electrons in highly charged ions. In this type of measurements, the mass differences of interesting candidates are typically well below $1$~u.

\section{Design of the PENTATRAP tower}\label{sec:Design}

\begin{figure*}
\centering
\includegraphics[width=0.9\textwidth]{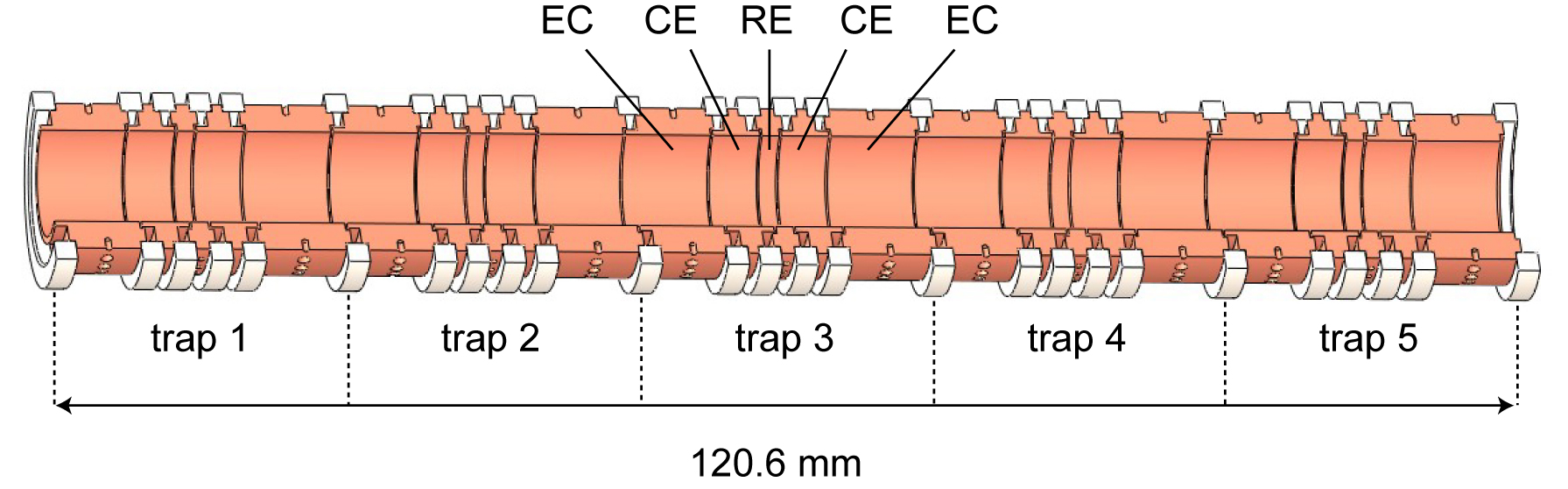}
\caption{Technical drawing of the complete trap tower consisting of five identical, cylindrical and compensated Penning traps. The electrodes are labeled RE for the ring electrode, CE for the correction electrodes and EC for the grounded endcaps. The length scales of the electrodes can be found in Tab.~\ref{tab:coefficients}. The electrodes are separated by sapphire rings, shown in white.}
\label{fig:falle}
\end{figure*}

The design of our trap tower consisting of five identical five-electrode traps is based on the potential given in Eq.~(\ref{eq:FallenPotential}). The Taylor coefficients of this potential were calculated using Eq.~(\ref{eq:Ckoeffizienten}). Since the required order of Bessel functions $n$ strongly increases with the total length and, therefore, the number of electrodes, all calculations were done with only three adjacent traps. 
For three traps the result of the calculations in the central trap are not affected by adding another electrode on the outside. Therefore, more electrodes are not needed for the calculations.
Furthermore, this shows that the approximation of the potential in Eq.~(\ref{eq:FallenPotential}), which vanishes at the outer ends of the trap tower (see discussion in Sec.~\ref{sec:CylPennPotential}), has no influence on the coefficients in the center and is therefore justified for our calculations.\\
For the real trap tower, the situation of two neighboring traps is common to the three inner traps and the results of the calculations are valid for them. But, certainly for the outer traps the performance will be slightly worse. On the other hand this is no issue to be worried about, since these traps are not dedicated for high-precision measurements (see \cite{Repp:11:ApplPhysB}).\\
Due to the linear dependence of the coefficients on all applied voltages, the performance of the central trap certainly depends to a small degree on the ring and correction voltages in the adjacent traps. Therefore, all calculations for the design of the trap were done with identical voltages in all three traps. This is the most common situation for the planned measurements since mainly mass doublets, which need almost identical trapping voltages, are foreseen to be measured. \\
To take advantage of the complete length of the homogeneous region of our superconducting magnet of $120$ mm, the length of $l=24$ mm per single trap was chosen.
From the discussion of the image charge shifts and the ion-ion interaction (Sec.~\ref{sec:IonIon}), an inner radius of $a=5$ mm was fixed as a compromise between these two effects. The length of the gaps between the electrodes was set to $d=150~\mu$m due to machining reasons.\\ 
In addition to finding an orthogonal geometry where $d_2$ vanishes, 
care was taken to keep the anharmonic shift in Eq.~(\ref{eq:AnharmonicShiftTaylor}) as low as possible in the final geometry. To this end, the trap should allow tuning out $c_4$ and the next anharmonic coefficient $c_6$ simultaneously. Therefore, the difference in tuning ratios $\Delta T\equiv \left.T\right|_{c_6=0}-\left.T\right|_{c_4=0}$ has to vanish.\\
In order to find a suitable geometry, using numerical methods the following three equations were solved simultaneously to find the unknown parameters $l_r$, $l_c$ and $T$:
\begin{eqnarray}
c_4(a, d, l_r, l_c, l, T)&=&0,\nonumber\\
c_6(a, d, l_r, l_c, l,T)&=&0,\nonumber\\
d_2(a, d, l_r, l,l_c)&=&0.
\end{eqnarray}
The numerical results are summarized in Tab.~\ref{tab:1}, where the geometrical lengths are rounded to $1~\mu$m for machining. In our case, the machining precision of the electrodes including a gold coating is limited to $\pm 5~\mu $m.  The electrostatic coefficients are calculated for the ideal geometry at the tuning ratio $T=\left.T\right|_{c_4=0}$. The errors resulting from limited machining precision are simply calculated by error 	propagation to estimate the worst case. In this respect, the contribution of the error of the gaps between the electrodes is approximately 10 times larger than the contribution from the error in any other dimensions of the trap. Note that the higher order coefficients still can be suppressed experimentally by finding a new tuning ratio which leads to $c_4=0$ in the real trap geometry. In Fig.~\ref{fig:falle}, a technical drawing of the resulting assembled trap tower is shown.

\begin{table}[h!]
\caption{Geometric and resulting electrostatic properties of a single trap within the PENTATRAP tower. The ideal geometry is found by minimizing $d_2$, $c_4$ and $c_6$ simultaneously. For machining, the geometric values are rounded to $1~\mu$m with respect to the ideal geometry. For an overview, the electrostatic coefficients up to $8$~-th order are calculated for the ideal geometric values at the tuning ratio of the ideal geometry $T=\left.T\right|_{c_4=0}$. Their errors are estimated for a machining precision of $\pm5~\mu$m. Note that the values of the higher order coefficients can still be suppressed experimentally by finding a new tuning ratio with respect to the real trap geometry.}
\label{tab:1}       
\begin{tabular}{l|r}
\hline\noalign{\smallskip}
$a$~~(mm)~~~~~~~~~~~~~~~~~~~~& 5 \\
\noalign{\smallskip}\hline\noalign{\smallskip}
 $d$~~(mm)& 0.15 \\
\noalign{\smallskip}\hline\noalign{\smallskip}
$l_r$~~(mm)& 1.457 \\
\noalign{\smallskip}\hline\noalign{\smallskip}
$l_c$~~(mm)& 3.932 \\
\noalign{\smallskip}\hline\noalign{\smallskip}
$l_e$~~(mm)& 7.040 \\
\noalign{\smallskip}\hline\noalign{\smallskip}
$\left.T\right|_{c_4=0}$& 0.881 \\
\noalign{\smallskip}\hline\noalign{\smallskip}
 $c_2~~\left(1/\text{mm}^2\right)$ & $(-1.496\pm0.007)\cdot10^{-2}$ \\
\noalign{\smallskip}\hline\noalign{\smallskip}
 $d_2~~\left(1/\text{mm}^2\right)$ & $0\pm1.241\cdot10^{-4}$ \\
\noalign{\smallskip}\hline\noalign{\smallskip}
 $c_4~~\left(1/\text{mm}^4\right)$ & $0\pm4.199\cdot10^{-6}$ \\
\noalign{\smallskip}\hline\noalign{\smallskip}
 $d_4~~\left(1/\text{mm}^4\right)$ & $(8.406\pm0.001)\cdot10^{-4}$ \\
\noalign{\smallskip}\hline\noalign{\smallskip}
 $c_6~~\left(1/\text{mm}^6\right)$ &  $0\pm1.892\cdot10^{-7}$  \\
\noalign{\smallskip}\hline\noalign{\smallskip} 
 $d_6~~\left(1/\text{mm}^6\right)$ & $(3.579\pm0.019)\cdot10^{-5}$  \\
\noalign{\smallskip}\hline\noalign{\smallskip} 
 $c_8~~\left(1/\text{mm}^8\right)$ &  $(1.672\pm0.077)\cdot10^{-7}$ \\
\noalign{\smallskip}\hline\noalign{\smallskip} 
 $d_8~~\left(1/\text{mm}^8\right)$ &  ~~$(-1.196\pm0.010)\cdot10^{-6}$\\
\noalign{\smallskip}\hline

\end{tabular}
\label{tab:coefficients}
\end{table}

\subsection{Orthogonality}
As a measure of the orthogonality of the trap the ratio $d_2/c_2$ can be taken, which is below $0.01$ in the worst case. Therefore, following from Eq.~(\ref{eq:AxialFreqIdeal}) and Eq.~(\ref{eq:CED}) a change of the tuning ratio shifts the axial frequency by only 
\begin{eqnarray}
\frac{\Delta \nu_z}{\Delta T}=\frac{1}{2}\frac{d_2}{c_2}\nu_z = \pm2.490~\frac{\text{Hz}}{\text{mUnit}}
\end{eqnarray}
at $\nu_z=600$~kHz, where the error is again determined by machining imperfections of $\pm 5~\mu $m. In this context, the expression mUnit simply denotes a change of $10^{-3}$ in the dimensionless tuning ratio $T$, which is a typical step size while tuning out anharmonicities.\\
Since typical narrow-band detection bandwidths are in the range of $10$ to $100$~Hz, the ion's axial frequency stays well in this range while tuning out anharmonicities in mUnit-steps of the tuning ratio. Therefore, the ion's signal is always visible in the tuning process, which is of practical importance. Furthermore, for uncorrelated changes in the voltages applied to the ring and compensation electrodes, the effect of $d_2$ can be expressed in relative terms using Eq.~(\ref{eq:AxialFreqIdeal}) and Eq.~(\ref{eq:CED}): 
\begin{eqnarray}
\frac{\delta\nu_z}{\nu_z}=\frac{1}{2}\sqrt{\left(\frac{e_2}{c_2}\frac{\delta U_0}{U_0}\right)^2+\left(\frac{d_2}{c_2}T\frac{\delta U_c}{U_c}\right)^2}.
\end{eqnarray}
Due to this, stability requirements of the correction voltage $U_c$ are at least 2 orders of magnitude less than for the ring voltage $U_0$.

\subsection{Compensation}
\begin{figure}
\centering
\includegraphics[width=0.45\textwidth]{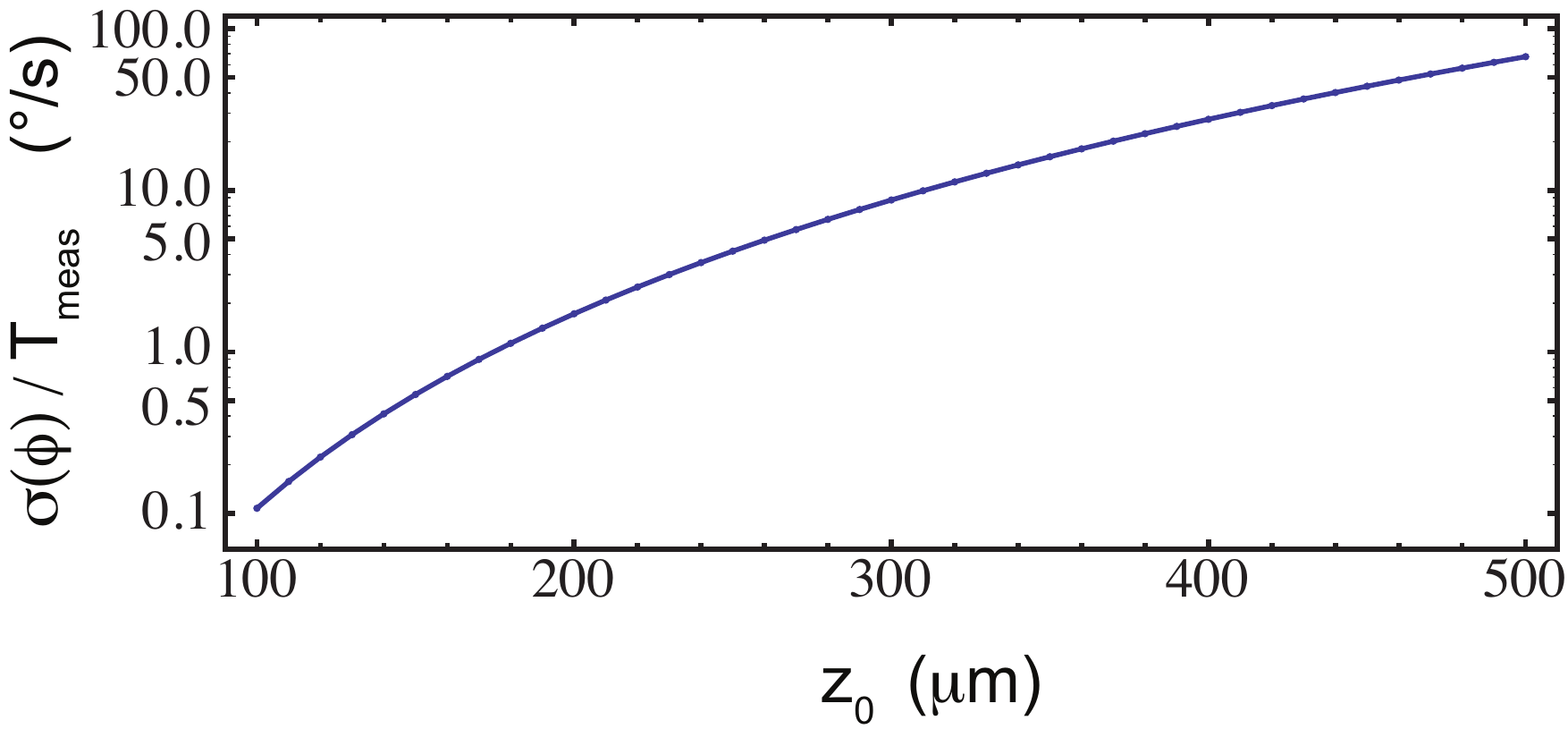}
\caption{Phase resolution $\sigma(\phi)$ due to anharmonic shifts in the axial oscillation frequency, normalized to the measurement time $T_{meas}$. The shifts are caused by a $10$~\% difference in the amplitude $z_0$ of two consecutive phase measurements (see text).}
\label{fig:phaseshifttime}
\end{figure}

In the PENTATRAP mass spectrometer we plan to use only cryogenic non-destructive ion detection. The axial frequency $\nu_z$ will be measured with the dip method \cite{Wineland:75:JApplPhys} at cryogenic temperatures of $4$~K. Due to the small particle amplitudes in the order of $10~\mu$m, anharmonic shifts are of no concern for this method. The same is true for the measurement of the magnetron frequency $\nu_-$, which is planned to be performed with the double-dip method explained in \cite{Verdu:04:PhysScripta}.\\
Anharmonic shifts are of more concern if detection methods with excited particle amplitudes are used. For our setup this is the case for the measurement of the radial frequency $\nu_+$. This concern is even more important, since the typical hierarchy of the three trap eigenfrequencies is $\nu_+\gg\nu_z\gg\nu_-$. Therefore, the accuracy achieved by mass measurements mainly depends on the accuracy of the measurement of $\nu_+$.\\
This frequency will be measured indirectly by a fast phase-sensitive detection method such as the so-called Pulse-aNd-Phase method (PNP) \cite{Cornell:90:PhysRevA} or a similar novel method \cite{Sturm:11:PRL}. In both cases, the phase of $\nu_+$ is transferred to the axial mode by side-band coupling, where it is read out by detection of the signal peak. The frequency of the modified cyclotron mode is then given by the derivative of its phase with respect to evolution time.\\
In general, the gain in measurement speed of phase-sensitive measurement methods compared to Fourier-limited frequency methods is approximately given by the factor $2\pi/\sigma(\phi)$, where $\sigma(\phi)$ is the resolution of the phase measurement \cite{Stahl:05:JPhysB}. For low amplitudes, the phase resolution is limited by thermal phase jitter resulting from the noise background of the detector. Therefore, as a premise for fast measurements for both methods, a sufficient signal-to-noise (SNR) ratio has to be ensured by exciting the ion to amplitudes far above the thermal noise floor. In this context, anharmonic shifts can result in an upper limit for the excitation of the amplitude. This limit will be roughly estimated for our trap in the following paragraph.\\
In most experimental situations it is convenient to tune out the leading anharmonic coefficient $c_4$ by choosing $T=\left.T\right|_{c_4=0}$. In this case, the remaining anharmonic shift is mainly given by the next anharmonic coefficient $c_6$. For the ideal geometry, both coefficients vanish for the same tuning ratio and the anharmonic shift given by Eq.~(\ref{eq:AnharmonicShiftTaylor}) therefore also vanishes.
For the real trap, this will not be the case anymore since these two coefficients vanish at different tuning ratios. However, it will still be possible to compensate $c_4$ only. To get an estimate of the remaining anharmonic shift in the real trap, a specific geometry has to be chosen since all coefficients $c_j(a, d, l_r, l_c, l, T)$ and therefore the tuning ratio $\left.T\right|_{c_4=0}=\left.T\right|_{c_4=0}(a, d, l_r, l_c, l)$ are functions of the real geometry.\\
As an example, we chose the rounded values for the trap geometry given in Tab.~\ref{tab:1}. In this case, the $6$-th order coefficient can be calculated to $c_6(\left.T\right|_{c_4=0}) \approx \pm 1.892\cdot 10^{-7}\cdot 1/\text{mm}^6 $ with worst case mechanical tolerances included, and the difference in the tuning ratio of $c_4$ and $c_6$ is given by $\Delta T\equiv \left.T\right|_{c_6=0}-\left.T\right|_{c_4=0}=22.9\cdot10^{-6}$. 
As a result, the remaining anharmonic shift at low energies given by Eq.~(\ref{eq:AnharmonicShiftTaylor}) ranges from $\partial^2 \nu_z/(\partial T\, \partial E_z)\approx 3.5$~mHz/(mUnit meV) to 
$1.4$~mHz/(mUnit meV) at $\nu_z=600$~kHz for masses of $100$~u to $240$~u, respectively.
As mentioned before, this will not be a problem for the low energy detection of $\nu_z$ and $\nu_-$.
However, for the phase detection of $\nu_+$ via the excited axial amplitude $z_0$ (see \cite{Cornell:90:PhysRevA,Sturm:11:PRL}), the following limitation arises:\\
 Due to finite reproducibility of the excitation pulse and different starting conditions in the cooled axial mode, the final excited axial amplitude $z_0$ differs between two consecutive measurements. A typical value reported to us is a $\sim 10$~\% deviation \cite{Sturm:11:PRL}. This in turn causes a shift of the frequencies of two consecutive measurements  due to the amplitude dependent anharmonic effects given by Eq.~(\ref{eq:AnharmonicShiftTaylor}). These frequency shifts limit the resolution of a phase measurement to $\sigma(\phi)=\Delta \omega_z\cdot T_{meas}$ \cite{Stahl:05:JPhysB}, where $T_{meas}$ is the measurement time for the axial motion. With our calculated value of $c_6$, this resolution normalized to the measurement time $T_{meas}$ is shown in Fig.~\ref{fig:phaseshifttime}.
For the measurement time, typically $\sim 3$ times the cooling time constant $\tau_z=\frac{D_{eff}^2}{R_0}\frac{m}{q^2}$ \cite{Wineland:75:JApplPhys} of the excited motion is used. This constant depends on the charge $q$ and mass $m$ of the ion. $R_0$ is the parallel resistance of the detection system described in \cite{Repp:11:ApplPhysB} for our case. $D_{eff}$ is the so-called effective electrode distance of the pickup electrode \cite{Stahl:98:phd}, which is a measure of the coupling strength of the detection electronics to the ion. In our case $D_{eff}$ is approximately $31.4$~mm for the endcap electrode, where the detection pickup is planned to be.\\
For charge states above $q=20+$ and our detection system, the expected cooling time constant can be estimated to be below $300$~ms even for masses up to $240$~u.\\
This implies, if a phase resolution of $10^\circ$ and therefore a gain in measurement speed of approximately one order of magnitude compared to frequency measurements is aimed for, we are limited to $300~\mu$m axial amplitude at a measurement time of $T_{meas} \sim 1$~s. On the other hand, this is of no concern since the signal strength scales with $q$ \cite{Wineland:75:JApplPhys} and therefore, at $q\geq20+$ amplitudes of $300~\mu$m should ensure more than sufficient SNR. \\

\subsection{Simultaneous measurement in adjacent traps}
The main advantage of our five-Penning-trap setup is the possibility to measure the cyclotron frequencies of two ions of different species in adjacent traps at the same time, and hence suppress the effect of magnetic field variation on the ratio. But, due to different charge-to-mass ratios there are different trapping voltages needed for the individual ions while the electrostatics of the traps was designed for identical voltages (Sec.~\ref{sec:Design}). Certainly, deviations from identical conditions can affect the performance due to the close distance and large openings of the traps. According to Eq.~(\ref{eq:AxialFreqIdeal}), the difference in the trapping voltages is determined by the difference in the $q/m$ ratio between the ions in two traps.\\
 In many cases, e.g. the measurement of \textit{Q}-values of $\beta$-transitions, which are interesting for the determination of the neutrino mass, the corresponding mass difference is very small. E.g. in the transition of $^{187}$Re$\rightarrow\,^{187}$Os, the \textit{Q}-value is only about $2.5$~keV. Thus, for the same charge state of the ions, almost identical trapping voltages are needed which differ  only by $\sim 0.7~\mu$V in this case.\\
The most critical case in this respect are measurements concerning binding energies of electrons in highly charged ions in order to test predictions of bound state QED theory. In this case the difference in the $q/m$ ratio of two charge states is much larger. For the measurement of the mass-ratio between Pb$^{81+}$ and Pb$^{82+}$ at $q/m = 0.39$, the difference is about $\Delta (q/m)\approx 1.2~\%$. The trapping voltages at $\nu_z=600$~kHz differ by $\Delta U\approx 0.15$~V. At this voltage difference between the traps, the anharmonic coefficients in both traps are about twice as large compared to the case of identical trapping voltages. Resulting from the discussion of anharmonic effects in the previous section, this will not limit our measurement resolution for high charge states of the ions.

\section{Summary}
In this article the design of a novel five-Penning-trap tower for the mass spectrometer PENTATRAP \cite{Repp:11:ApplPhysB} was presented. The potential of a five-electrode cylindrical Penning trap inside a trap tower was calculated from the Laplace-equation and the electrostatic properties analyzed in terms of a Taylor expansion of the potential. Special attention was payed to the orthogonality of the calculated trap geometry as well as the possibility to tune out the leading two anharmonic coefficients of the electrostatic potential simultaneously. 
The experimental limitations due to anharmonic effects at large amplitudes were estimated, which is important for phase-sensitive measurement methods of the reduced cyclotron frequency. Furthermore, the systematic effects of highly charged ions in terms of image charge shifts of the free cyclotron frequency as well as Coulomb interaction of two ions in adjacent traps were estimated for the resulting trap geometry given in Tab.~\ref{tab:coefficients}. \\\\
This work is supported by the Max-Planck Society and by the Deutsche Forschungsgemeinschaft under contract BL 981/2-1. Yu.~N.~and Ch.~B.~thank the support from the Extreme Matter Institute (EMMI). S.~U.~acknowledges support from the IMPRS-QD. C.~R. thank M. Kretzschmar and U. Warring for fruitful discussions and support.

%

\begin{thebibliography}{}
%
%
\bibitem{Blaum:10:ContPhys}
K. Blaum, Yu. Novikov, G. Werth, 
Contemp. Phys. \textbf{51}, (2010) 149-175.

\bibitem{Blaum:06:PhysRep}
K. Blaum, 
Phys. Rep. \textbf{425}, (2006) 1-78.

\bibitem{Bergstroem:03:EurPhyJD}
I. Bergstr\"om, M. Bj\"orkhage, K. Blaum, H. Bluhme, T. Fritioff, Sz. Nagy, R. Schuch, 
Eur. Phys. J. D \textbf{22}, (2003) 41-45.

\bibitem{Nagy:06:PhysRevLett}
Sz. Nagy, T. Fritioff, M. Suhonen, R. Schuch, K. Blaum, M. Bj\"orkhage, I. Bergstr\"om, 
Phys. Rev. Lett. \textbf{96}, (2006) 163004.

\bibitem{Smith:08:PhysRevLett}
M. Smith, M. Brodeur, T. Brunner, S. Ettenauer, A. Lapierre, R. Ringle, V. L. Ryjkov, F. Ames, P. Bricault, G. W. F. Drake, P. Delheij, D. Lunney, F. Sarazin, J. Dilling, 
Phys. Rev. Lett. \textbf{101}, (2008) 202501.

\bibitem{Block:10:Nature}
M. Block, D. Ackermann, K. Blaum, C. Droese, M. Dworschak, S. Eliseev, T. Fleckenstein, E. Haettner, F. Herfurth, F. P. He\ss berger, S. Hofmann, J. Ketelaer, J. Ketter, H.-J. Kluge, G. Marx, M. Mazzocco, Yu. Novikov, W. R. Pla\ss , A. Popeko, S. Rahaman, D. Rodriguez, C. Scheidenberger, L. Schweikhard, P. G. Thirolf, G. K. Vorobyev, C. Weber, 
Nature \textbf{463}, (2010) 785-788.

\bibitem{Rainville:05:Nature}
S. Rainville, J. K. Thompson, E. G. Myers, J. M. Brown, M. S. Dewey, E. G. Kessler, R. D. Deslattes, H. G. B\"orner, M. Jentschel, P. Mutti, D. E. Pritchard, 
Nature \textbf{438}, (2005) 1096-1097.

\bibitem{VanDyck:04:PhysRevLett}
R. S. Van Dyck, S. L. Zafonte, S. Van Liew, D. B. Pinegar, P. B. Schwinberg, 
Phys. Rev. Lett \textbf{92}, (2004) 220802.

\bibitem{Brown:82:PhyRevA}
L. S. Brown, G. Gabrielse, 
Phys. Rev. A \textbf{25}, (1982) 2423-2425.

\bibitem{Brown:86:RevModPhys}
L. S. Brown, G. Gabrielse, 
Rev. Mod. Phys. \textbf{58}, (1986) 233-311.

\bibitem{Kellerbauer:02:EurPhyJD}
A. Kellerbauer, K. Blaum, G. Bollen, F. Herfurth, H.-J. Kluge, M. Kuckein, E. Sauvan, C. Scheidenberger, L. Schweikhard, 
Eur. Phys. J. D \textbf{22}, (2002) 53-64.

\bibitem{VanDyck:06:IntJMS}
R. S. Van Dyck, D. B. Pinegar, S. Van Liew, S. L. Zafonte, 
Int. J. Mass Spectrom. \textbf{251}, (2006) 231-242.

\bibitem{VanDyck:99:RevScI}
R. S. Van Dyck, D. L. Farnham, S. L. Zafonte, P. B. Schwinberg, 
Rev. Sci. Instrum. \textbf{70}, (1999) 1665-1671.

\bibitem{Rainville:04:Science}
S. Rainville, J. K. Thompson, D. E. Pritchard, 
Science \textbf{303}, (2004) 334-338.

\bibitem{Repp:11:ApplPhysB}
J. Repp, Ch. B\"ohm, J. R. Crespo Lopez-Urrutia, A. D\"orr, S. Eliseev, S. George, M. Goncharov, Yu. Novikov, C. Roux, S. Sturm, S. Ulmer, K. Blaum, preceding article in this issue.


\bibitem{Gabrielse:84:IntJMS}
G. Gabrielse, F. C. MacKintosh, 
Int. J. Mass Spectrom. \textbf{57}, (1984) 1-17.



\bibitem{Blaum:09:JPhysB}
K. Blaum, H. Kracke, S. Kreim, A. Mooser, C. Mrozik, W. Quint, C. C. Rodegheri, B. Schabinger, S. Sturm, S. Ulmer, A. Wagner, J. Walz, G. Werth, 
J. Phys. B \textbf{42}, (2009) 154021.

\bibitem{Hanneke:08:PhysRevLett}
D. Hanneke, S. Fogwell, G. Gabrielse, 
Phys. Rev. Lett. \textbf{100}, (2008) 120801.

\bibitem{Kluge:08:AdvQuanChem}
H.-J. Kluge, T. Beier, K. Blaum, L. Dahl, S. Eliseev, F. Herfurth, B. Hofmann, O. Kester, S. Koszudowski, C. Kozhuharov, G. Maero, W. N\"ortersh\"auser, J. Pfister, W. Quint, U. Ratzinger, A. Schempp, R. Schuch, Th. St\"ohlker, R. C. Thompson, M. Vogel, G. Vorobjev, D. F. A. Winters, G. Werth, 
Adv. Quantum Chem. \textbf{53}, (2008) 83-98.



\bibitem{Gabrielse:86:PhysRevLett}
G. Gabrielse, X. Fei, K. Helmerson, S. L. Rolston, R. Tjoelker, T. A. Trainor, H. Kalinowsky, J. Haas, W. Kells, 
Phys. Rev. Lett. \textbf{57}, (1986) 2504-2507.

\bibitem{Gabrielse:02:PhysRevLett}
G. Gabrielse, N. S. Bowden, P. Oxley, A. Speck, C. H. Storry, J. N. Tan, M. Wessels, D. Grzonka, W. Oelert, G. Schepers, T. Sefzick, J. Walz, H. Pittner, T. W. H\"ansch, E. A. Hessels,
Phys. Rev. Lett. \textbf{89}, (2002) 213401.

\bibitem{Amoretti:02:Nature}
M. Amoretti, C. Amsler, G. Bonomi, A. Bouchta, P. Bowe, C. Carraro, C. L. Cesar, M. Charlton, M. J. T. Collier, M. Doser, V. Filippini, K. S. Fine, A. Fontana, M. C. Fujiwara, R. Funakoshi, P. Genova, J. S. Hangst, R. S. Hayano, M. H. Holzscheiter, L. V. Jorgensen, V. Lagomarsino, R. Landua, D. Lindel\"of, E. Lodi Rizzini, M. Macri, N. Madsen, G. Manuzio, M. Marchesotti, P. Montagna, H. Pruys, C. Regenfus, P. Riedler, J. Rochet, A. Rotondi, G. Rouleau, G. Testera, A. Variola, T. L. Watson, D. P. van der Werf, 
Nature \textbf{419}, (2002) 456-459.

\bibitem{VanDyck:76:ApplPhysLett}
R. S. Van Dyck, D. J. Wineland, P. A. Ekstrom, H. G. Dehmelt, 
Appl. Phys. Lett. \textbf{28}, (1976) 446-448.

\bibitem{jacksonPot}
J. D. Jackson, \textit{Classical Electrodynamics} (John Wiley \& Sons, 1975 2nd ed.)  108.

\bibitem{Gabrielse:89:IntJMS}
G. Gabrielse, L. Haarsma, S. L. Rolston, 
Int. J. Mass Spectrom. \textbf{88}, (1989) 319-332.

\bibitem{Verdu:08:NJP}
J. Verdu, S. Kreim, K. Blaum, H. Kracke, W. Quint, S. Ulmer, J. Walz, 
New J. Phys. \textbf{10}, (2008) 103009.


\bibitem{Gabrielse:83:PhysRevA}
G. Gabrielse, 
Phys. Rev. A \textbf{27}, (1983) 2277-2290.


\bibitem{jackson}
J. D. Jackson, \textit{Classical Electrodynamics} (John Wiley \& Sons, 1975 2nd ed.)  134.

\bibitem{VanDyck:89:PhyRevA}
R. S. Van Dyck, F. L. Moore, D. L. Farnham, P. B. Schwinberg, Phys. Rev. A \textbf{40,} (1989) 6308-6313.

\bibitem{Porto:01:PhysRevA}
J. V. Porto, 
Phys. Rev. A \textbf{64}, (2001) 023403.

\bibitem{Haeffner:00:phd}
H. H\"affner, 
\textit{Phd Thesis} (Universit\"at Mainz, 2000).

\bibitem{Wineland:75:JApplPhys}
D.J. Wineland, H.G. Dehmelt, 
J. Appl. Phys. \textbf{46}, (1975) 919-930.

\bibitem{Verdu:04:PhysScripta}
J. Verdu, S. Djekic, S. Stahl, T. Valenzuela, M. Vogel, G. Werth, H.-J. Kluge, W. Quint, 
Phys. Scripta \textbf{T112}, (2004) 68-72.

\bibitem{Cornell:90:PhysRevA}
E. A. Cornell, R. M. Weisskoff, K. R. Boyce, D. E. Pritchard, 
Phys. Rev. A \textbf{41}, (1990) 312-315.

\bibitem{Sturm:11:PRL}
S. Sturm, B. Schabinger, A. Wagner, K. Blaum, accepted by Phys. Rev. Lett.  (2011)

\bibitem{Stahl:05:JPhysB}
S. Stahl, J. Alonso, S. Djekic, H.-J. Kluge, W. Quint, J. Verdu, M. Vogel, G. Werth, 
J. Phys. B \textbf{38}, (2005) 297-304.

\bibitem{Sturm:11:private}
S. Sturm, private communication

\bibitem{Jefferts:93:RevSciInstr}
S. R. Jefferts, T. Heavner, P. Hayes, G. H. Dunn, 
Rev. Sci. Instrum. \textbf{64}, (1993) 737-740.

\bibitem{Stahl:98:phd}
S. Stahl, 
\textit{Phd Thesis} (Universit\"at Mainz, 1998).

\end{thebibliography}
%


\end{document}